**Title**
Navigation between initial and desired community states using shortcuts

**Running title**
Navigation between community states

**Type of article**
Letter

**Authors**
Benjamin W. Blonder 1, *, +
Michael H. Lim 2, *
Zachary Sunberg 3
Claire Tomlin 2

**Affiliations**
1: Department of Environmental Science, Policy, and Management, University of California Berkeley, Berkeley, CA, USA
2: Department of Electrical Engineering and Computer Science, University of California Berkeley, Berkeley, CA, USA
3: Aerospace Engineering Sciences Department, University of Colorado Boulder, Boulder, CO, USA
*: these authors contributed equally
+: Corresponding author: email benjamin.blonder@berkeley.edu; mailing address 54 Mulford Hall, UC Berkeley, Berkeley, 94720-3114, USA

**Number of**
Words in abstract: 149; Words in main text: 4963; Number of references: 54; Number of figures: 5; Number of tables: 1

**Abstract**

Ecological management problems often involve navigating from an initial to a desired community state. We ask whether navigation is possible without brute-force additions and deletions of species, using actions of varying costs: adding/deleting a small number of individuals of a species, changing the environment, and waiting. Navigation can yield direct paths (single sequence of actions) or shortcut paths (multiple sequences of actions with lower cost than a direct path). We ask (1) when is non-brute-force navigation possible?; (2) do shortcuts exist and what are their properties?; and (3) what heuristics predict shortcut existence? Using several empirical datasets, we show that (1) non-brute-force navigation is only possible between some state pairs, (2) shortcuts exist between many state pairs; and (3) changes in abundance and richness are the strongest predictors of shortcut existence, independent of dataset and algorithm choices. State diagrams thus unveil hidden strategies for efficiently shifting between states.

**Introduction**

Many ecological management problems involve observing a community in an initial state, then taking a sequence of actions to yield a desired state (e.g. promoting gut microbiome health after infection, restoring a degraded rangeland). Management problems are often solved by brute-force navigation, which involves removing all individuals of undesired species and adding many individuals of desired species at great effort (e.g., antibiotics+probiotics, bulldozing+replanting). Such navigation may succeed, but at high cost and impact. Alternatives may exist that are more efficient and have fewer side effects. The challenge is to identify action sequences, i.e. navigation, that yield the desired state at lower cost and effort.

Some prior navigation approaches focused on continuous control of community *dynamics*. The problem has recently been conceptually explored in models (Angulo *et al.* 2019; Jones *et al.* 2020; Brias & Munch 2021; Baranwal *et al.* 2022). Applications exist for e.g., fisheries, forestry, agriculture, and other natural resource management challenges where continual intervention is of interest (Krausman *et al.* 2013; Boettiger *et al.* 2015; Palmer *et al.* 2016; Lapeyrolerie *et al.* 2022), and also in microbial systems where metabolite production or infectious disease are priorities (Costello *et al.* 2012; García-Jiménez *et al.* 2018; Angulo *et al.* 2019). However, continuous control of multiple species' abundances becomes mathematically prohibitive and biologically unrealistic in high-richness communities.

We instead propose a discretized navigation approach that focuses on control of community *outcomes*. Many management problems can be simplified to coexistence outcomes (Maynard *et al.* 2020; Clark *et al.* 2021; Blonder & Godoy 2022). Reaching an outcome (desired state) may be more important than the transient dynamics. Formulation as a discrete path planning problem can reduce mathematical complexity and improve biological realism.

Our hypothesis is that the internal dynamics of a community enable taking actions that nudge a community between states, either through direct paths or shortcut paths, both of which are lower-effort than brute-force navigation. Here, we define a direct path as a single set of low-effort actions that yield the desired state, and a shortcut path as a sequence of multiple sets of low-effort actions that yield the desired state; action sets are separated by waiting for the community to reach a feasible and stable fixed point. The term 'shortcut' is used to indicate that path *cost* is small, not path *length* (**Figure 1a**). Conceptually, we nudge a community until it tips into an alternate basin of attraction, then repeat this nudging process until the desired state is reached. Several small nudges may be lower in cost than a single large push into the desired state.

Brute-force navigation is always theoretically possible between states by removing all individuals of undesired species and then adding a sufficient number of desired species, ignoring the internal dynamics of the community. However, brute-force is often impractical, and if the desired state is not feasible and stable, further continuous effort would be needed to maintain the state. We focus therefore on finding direct and shortcut paths between feasible and stable states only. Direct paths may be findable via trial-and-error. However, shortcuts are difficult to find because of the near-infinite numbers of potential action sequences to explore.

The navigation problem is loosely analogous to the game of 'Snakes and Ladders' (known originally as 'gyān caupaṛ') (**Figure 1b**) (Topsfield 2006). In this game, "*the player should complete [the tour of] the board according to its numbering, starting at birth and ending at liberation. Going upward comes about by means of the ladder; going down comes about from the body of the snake. Going up is achieved from good actions; [going down from] the face of the snake is caused by bad actions. Vaikuṇṭha [the heaven of Viṣṇu] is reached by completing the* *game; otherwise the player must go on climbing*" (Harikrishna 1871). While Snakes and Ladders is a game of chance, not choice, our hypothesis is approximately equivalent to finding and then using 'snakes' (richness-decreasing shortcuts) and 'ladders' (richness-increasing shortcuts) to navigate between states.

First, we show how to enumerate a state diagram characterizing all of the possible transitions between all possible fixed point states. We consider actions that include adding $\epsilon$ individuals of a certain species, deleting $\epsilon$ individuals of a certain species, changing the environment, or waiting. The first three actions are assumed to occur instantaneously, shifting the community into a transient state, while the last action takes time, shifting the community to a fixed point. Each type of action $i$ is also assumed to have a different cost $C_i$. We then show how to identify shortcuts on the state diagram for arbitrary pairs of initial and desired states.

We apply the approach to six empirical parameterizations of the generalized Lotka-Volterra model, varying in taxonomy and species pool richness. We use these data to ask: (1) when is navigation between states possible without using brute-force; (2) are shortcut paths common, and what are their characteristics; and (3) are shortcuts predictable based on dataset or initial/desired community properties?

**Materials and Methods**

<u>The state diagram approach</u>

There is a set of *n* species comprising a regional pool, of which any subset may co-occur locally in the community. The state of the community, $X = \{X_i(t)\} \in \Re^n_{\geq 0}$, is defined as the vector of abundances of each species $X_i$ (1≤*i*≤*n*) at a time *t*. There is a set of discrete environments with cardinality *m* defined by $\{E_j\} \in M$ with $1 \leq j \leq m$. Note that the environment may actually be continuous; here we simply consider some discrete points within the environment to be reachable by actions, e.g., to model cases where an experimentalist could select among 'cold' to 'warm' and 'hot' conditions (*m*=3). There is a dynamical model that predicts temporal changes in the state as a function of variables, which may include *X* and *E*, $\frac{dX(t)}{dt} = f(X(t), E)$.

Based on this dynamical model, there are a set of fixed points with cardinality $\Xi$, $\{\xi_k\}$ with $1 \leq k \leq \Xi$, defining the points where $\frac{dX(t)}{dt} = 0$. Note that if *E* changes, so too may $\Xi$. A fixed point *k* can have an attribute $fs(\xi_k)$ indicating that it is feasible (i.e. all species present *i* occur at non-negative abundances; $\xi_{k,i} \geq 0$) and stable (for every small $\varepsilon > 0$ there exists a $\delta > 0$ such that if $|X(t_0) - \xi_k| < \delta$ then $|X(t) - \xi_k| < \varepsilon$ for all $t \geq t_0$).

We next enumerate $\{\xi_k\}$ over all combinations of species being present or absent in the community (i.e. the empty community, all species occurring alone, all pairs, all triplets, etc.). These fixed points can be identified by exploring every subspace of the state space (all combinations of presences/absences), then re-calculating dynamical model nullclines.

We consider four types of actions, indexed $1 \leq q \leq 4$. Each action type *q* is assumed to have some cost $C_q \geq 0$ and have a different consequence: (*q*=1) adding a small number ($\epsilon$) of individuals of a single species *i* (i.e. $X_i \to X_i + \epsilon$); (*q*=2) deleting a small number ($\epsilon$) of individuals of a single species *i* (i.e. $X_i \to max(X_i - \epsilon, 0)$); (*q*=3) changing the state of the environment *j* to *j**, $1 \leq j^* \leq m$ ($E_j \to E_{j^*}$, no change to *X*), and (*q*=4) waiting for a shift into fixed point *k**, $1 \leq k^* \leq \Xi$ ($X \to \xi_{k^*}$, no change to *E*). Each species *i* can either be added or deleted up to one time until a waiting action has been performed. For all actions except waiting, the consequence is assumed to occur instantaneously; for waiting, the consequence is assumed to occur as $t \to \infty$ and determined by the dynamical model. That is, we assume that states do not reach a fixed point until a waiting action, and that multiple non-waiting actions can be taken in sequence before waiting.

The system can now be discretized into a smaller state space $\{Y\}$ that describes fixed points and transient points. In each environment $E$, we therefore assume that each state can either be at one of the fixed points $\xi_k$ or, for each fixed point, at one of the $3^n$ possible transient $\epsilon$-addition or $\epsilon$-deletion states that occur immediately after any number of actions is taken. The overall cardinality of the discretized state space $\{Y\}$ is therefore $m \times \Xi \times 3^n$ or approximately $m \times 2^n \times 3^n$ assuming one fixed point per species combination. The action space can also be discretized. There are a total of $n$ $\epsilon$-additions and $\epsilon$-deletions, $m$ environmental changes, and 1 wait action, yielding a cardinality of $2n + m + 1$. Each action, now by definition, yields a transition from a state in $\{Y\}$ to another state in $\{Y\}$.

With this information for fixed points and the outcomes of actions at each fixed point, we can construct a directed graph called the state diagram. Vertices are states in $\{Y\}$ and edges are actions, where the arrow head is the state after the action and the arrow base is the state before the action. Each vertex $k$ has attribute $fs(\xi_k)$; each edge $\delta$ has an attribute $C_q$. We define an action sequence $\Delta = \{\delta_1, \delta_2, ...\}$ as an ordered set of edges (actions) that connects an initial vertex (state) to a desired vertex (state), with associated cost sequence $\omega = \{C_{q,1}, C_{q,2}, ...\}$.

Our primary insight is that the navigation problem is now equivalent to a shortest-path (lowest-cost) problem on a directed graph (the state diagram), i.e. finding a Δ that minimizes $\sum \omega$. This general mathematical problem can be solved efficiently (Ford Jr 1956; Cherkassky et al. 1996). If a path does not exist, the only solution is brute-force; if a path does exist, and has one wait action, it is direct, and if it has more than one wait action, it is a shortcut.

Implementation

We implemented the state diagram approach for the GLV model, which has been widely studied to explore questions of species coexistence (Barabás *et al.* 2016; Saavedra *et al.* 2017) and can accommodate cases where the environment influences parameter values (Van Dyke *et al.* 2022). The dynamical model is

$$\frac{dX(t)}{dt} = diag(X(t))(r(E) + A(E)X(t))$$

where $E$ is assumed constant unless changed by an action. Here, $r(E)$ is a $n \times 1$ vector that indicates the intrinsic growth rates of each species, and $A(E)$ is a $n \times n$ matrix whose *i,j* entry represents the change in species *i*'s per-capita growth rate for a unit change in the density of species *j*.

If *A* is non-singular, for each parameter combination, there is one non-trivial fixed point, determinable by nullcline analysis:

$$\xi = -A^{-1}(E)\, r(E)$$

If the fixed point is not feasible, the state will shift to a subspace with some species absent (see below). If *A* is singular, there can be many fixed points corresponding to the null space of *A*, corresponding to cases where parameters are either linear combinations or there is partitioning in the interaction network (Angulo *et al.* 2019). Stability is defined by the criterion

$$max_i[Re(\{\lambda_i(E)\})] < 0$$

where $\{\lambda_i(E)\}$ are the *n* eigenvalues of *A(E)*, and feasibility is determined based on the values of ξ.

To then calculate all the fixed points Ξ, the process can be iterated for all combinations of species. Because all GLV interactions are pairwise, cases where species *j* is absent can be handled by dropping row *j* and column *j* of the *A* matrix (i.e. obtaining the principal submatrix), and simultaneously dropping entry *j* of the r vector. Multiple entries can be dropped in cases where multiple species are absent. This is non-trivial, as the eigenvalues of a principal submatrix (which are closely related to the matrix inverse, and thus the location of the fixed point) are not necessarily the same as for the original matrix (Johnson & Robinson 1981). That is, combinations of species may behave differently from subsets of those combinations (Saavedra *et al.* 2017), a phenomenon also seen in models with higher-order interactions (Mayfield & Stouffer 2017). If *A* and all its principal submatrices are non-singular, then there is a single fixed point per iteration, yielding $\Xi = 2^n$ fixed points for each value of *E*. If *A* is singular, there may be more or fewer fixed points to be considered.

The outcomes of actions are determined based on numerical integration of the dynamical model. First, we enumerate all desired addition, deletion and environment change actions for each fixed point, arriving at intermediate states. Then, when the waiting action is performed, the initial condition of the numerical integration is set to the intermediate state, and the dynamics are run forward with integration time span proportional to the smallest eigenvalue of *A* to ensure that the system can approach equilibrium. The resulting final abundances are then matched to the corresponding fixed point if the integration is successful and results in a non-trivial fixed point.

We calculate Δ and ∑ω for pairs of starting and desired states using A* search, which is a best-first search algorithm that expands local paths around the source vertex according to a combination of the cost of the path from the initial vertex to the current vertex plus the cost of a heuristic estimate of the cost from the current vertex to the desired vertex. It is guaranteed to find a solution if one exists (Hart *et al.* 1968). We use an admissible heuristic that optimistically assumes that a single round of adding small numbers of individuals of currently missing species followed by a waiting action is sufficient to reach the desired basin of attraction. All algorithms were implemented in Julia (version 1.6.0). ODEs were solved using Rodas4P with absolute tolerance $10^{-6}$, relative tolerance $10^{-6}$, and max iterations $10^{3}$.

Empirical parameterization

We studied six cases where parameter estimates for *A* and *r* come from fitting generalized Lotka Volterra models to empirical data (**Table 1**, taxon names in **Table S1**). These comprise: ('Ciliate') a *n*=5 protozoan ciliate community (Maynard *et al.* 2020) based on data for 19 °C growth; ('Ciliate+environment3') the above *n*=5 community for *m*=3 environments: 15, 19, and 23 °C growth from (Pennekamp *et al.* 2018), ('Ciliate+environment5') as above for *m*=5 environments also including 17 and 21°C growth; ('Human gut') a *n*=12 *m*=1 synthetic human gut microbial community (Venturelli *et al.* 2018); ('Mouse gut') a *n*=11 *m*=1 mouse gut microbial community including the difficult-to-remove pathogen *Clostridium difficile* (Stein *et al.* 2013) based on data from (Buffie *et al.* 2012); and ('Protist') a *n*=11, *m*=1 protist and rotifer community based on *A* values from (Carrara *et al.* 2015) and *r* values from (Carrara *et al.* 2012) and supplemented by additional *r* values for two missing taxa (pers. comm. F. Altermatt, May 7, 2021).

Computational experiments

We performed A* experiments over all multiple action cost combinations and action magnitudes. Addition and deletion actions used $\epsilon$ in $\{10^{-1}, 10^{-3}, 10^{-5}\}$. Each type of action *q* used costs in $\{10^{-1}, 10^{0}, 10^{1}\}$. We also tested whether capping the total number of actions before a wait (a scenario where actions should be simple to implement) influenced navigation. This comprises $3 \times 3^4 \times 2 = 486$ experiments per dataset. Impacts of capping were minimal so main-text results only consider no capping, with capped results provided in output files. For each dataset, we picked 10,000 random pairs of initial and desired states. We determined whether a non-brute-force navigation solution existed for each dataset for 10,000 subsampled state pairs for which both start and end states are feasible and stable. State pairs were sampled without replacement using a fixed random number generator seed within each dataset to enable direct comparison between experimental results with different hyperparameter choices.

Statistical analysis

To address Question 1, for each A* experiment, we determined whether non-brute-force navigation was possible via any path. We also visualized state diagrams for selected cases, and determined whether, across cases, some intermediate states were more commonly visited (i.e. variation in node degree and centrality).

To address Question 2, for each A* experiment where non-brute-force navigation was possible, we determined whether the lowest-cost path was direct or a shortcut. We assessed variation in path length, and also visualized paths for selected cases.

To address Question 3, we built a random forest model that outputs probabilities, where path type (brute-force, direct, shortcut) was the dependent variable. Predictor variables reflected several easily-measured state properties, assuming no knowledge about the state diagram or the GLV dynamics: change in mean abundance, richness, and Jaccard similarity between initial and desired states; $log_{10}\epsilon$; $n$; $m$; all four costs $C_q$; and dataset name. To reduce computational costs, a subset of 100 (or the maximum available) state pairs were randomly sampled from each of the 2916 A* experiments, after which we balanced the sampling by path type (brute-force, direct, shortcut) to the minimum number of samples available in each type. The final dataset comprised 25,782 cases. We used default parameters in the *ranger* package (version 0.14.1). We calculated a permutation importance for each predictor, made partial dependence plots for the most important predictors, and calculated overall accuracy using a 10-fold cross-validation in the *caret* package (version 6.0-93). All analyses were performed in R (version 4.2.0).

**Results**

Question 1: Navigation

State diagrams had complex topologies that varied widely with dataset (**Figure 2**). Some datasets only contained a small fraction of feasible and stable states, limiting non-brute-force navigation among low richness states (e.g., protist), while others supported navigation to high richness states (e.g., mouse gut). Higher-richness transient states used for navigation occurred widely in all datasets, indicating that species interactions, here competitive exclusions, played a key role in navigation, but also represented a potential hazard if they would be unsafe to reach (see Discussion). Actions were dominated by additions in some datasets (e.g., human gut, mouse gut) and by deletions in others (e.g., ciliate+environment3, ciliate+environment5), though actions comprising both additions/deletions also occurred (**Figure S1**).

Varying the GLV parameterization influenced the state properties, and thus the possible navigation targets. Varying $\epsilon$ changed the topology of the state diagram, with larger $\epsilon$ often resulting in more edges, but sometimes loss of edges (**Figure S2**). For a fixed state diagram topology, varying the costs $C_q$ influenced the edge weights and thus the navigation paths.

Navigation probabilities, defined as the number of state pairs connected by a non-brute-force path divided by the number of feasible and stable state pairs, varied widely (**Figure 3a**). Probabilities were lowest for the human gut and highest for the ciliate+environment5 dataset. Increasing $\epsilon$ increased probabilities for all datasets. Some intermediate states were consistently visited (**Figure S3**), showing that there are hubs on the state diagram. However, hubs were not common in the ciliate+environment datasets, suggesting that environmental variation enables more diverse navigation pathways. Hubs were not correlated with in-degree or out-degree on the state diagram (**Figure S4**). In general, there was a tradeoff between in- and out-degree, indicating that states that are easier to reach are harder to leave, and vice versa.

Question 2: Shortcut properties

Shortcut probabilities, defined as the probability a state pair was connected by a shortcut, conditioned on navigation between the states being possible, also varied substantially (**Figure**

**3b**). Shortcut probabilities ranged from 14% to 71% across datasets and $\epsilon$ values, except for the ciliate dataset at 0%. Increasing $\epsilon$ did not consistently increase shortcut probability.

Among shortcut paths, the number of steps varied widely (**Figure S5**). The mouse gut and ciliate+environment datasets consistently had the longest path lengths, some involving as many as eight sequential actions, which is consistent with the greater number of links present in their state diagrams (**Figure 2**). Other datasets typically involved paths comprising 2-4 actions.

Visualizing shortcut paths illustrates the complexity of navigation. In the mouse gut, completely removing the pathogen *C. difficile* when it is initially present was often possible. For the experimental conditions described in **Figure 2**, we found 4,304/10,000 cases with the pathogen present; of these, a complete removal via shortcut was possible in 111 cases. Two examples are shown in Figure **4a-b**. Similarly, community turnover is often achievable by leveraging environmental change, as in the ciliate+environment5 dataset. Also for the experimental conditions described in **Figure 2**, we found 716/10,000 cases that had no *net* change from 15°C growth; of these, reduction in richness via shortcuts leveraging environmental change was possible in 206 cases. Two examples are shown in Figure **4c-d**. In all cases, navigation used timely actions to cause useful competitive exclusions, which allowed jumping between states until the desired state was reached. In other cases (not shown) where $C_{\epsilon-deletion}$ is assumed smaller, $\epsilon$-deletions were more commonly used.

Question 3: Predicting shortcuts

The random forest model of path type (brute-force, direct, shortcut) had a cross-validation accuracy of 77.2%. Permutation importances of predictors varied widely (**Figure S6**). The most important predictors were ΔRichness and ΔAbundance (desired state value minus initial state value) (**Figure 5, Figure S7**). Shortcut paths were most probable when ΔRichness was positive and ΔAbundance was negative, i.e. cases involving introducing species and displacing dominant species. Shortcut paths were also more probable when Jaccard similarity was small, $\epsilon$ was large, and *m* was large; costs had negligible effects (**Figure 5**).

**Discussion**

We showed that navigation between states is an equivalent problem to searching for lowest-cost sequences of actions that comprise direct and shortcut paths. Shortcuts can be obtained by using small sequential abundance perturbations (e.g. low-density introductions) and environment perturbations to nudge communities between states. Shortcuts were most probable when large richness-increasing, abundance-decreasing, similarity-decreasing state shifts were desired, when perturbation size ($\epsilon$) was large, and when environmental change was possible. Thus, our work suggests that brute-force approaches to navigation like antibiotics or clearcutting may have realistic and less impactful alternatives.

Application cases

The approach could be used for navigation problems where there are a finite number of fixed points to be considered, and where the time to reach a fixed point is substantially smaller than the timescale of the overall problem. Realistic application cases may include communities with fast population dynamics, e.g., microbial communities or bioreactors/chemostats, or annual plants. Optimistic application cases could include resolving human health problems that are linked to the microbiome, (Sonnenburg 2015; Sonnenburg & Sonnenburg 2019), e.g. *C. difficile* removal, or improvement of crop/soil health via associated microbial communities (Mueller & Sachs 2015). Additionally, applications could be possible in annual plant restoration projects (D'Antonio & Meyerson 2002; Perring *et al.* 2015).

The state diagram approach could also be useful for assembling synthetic communities, e.g. in microbial bioreactor applications (Clark *et al.* 2021; Baranwal *et al.* 2022). This problem maps onto the navigation problem, because the desired state is a certain feasible and stable community and the initial state is an empty community. Action sequences could be identified to achieve these goals when brute-force assembly of the desired state is not possible or efficient.

Extensions to the navigation approach

We implicitly assumed that the species pool richness was relatively low, which allowed us to use the A* algorithm. This algorithm does not work well when $n$ or $m$ are large, because the state diagram becomes too large to explore. However, the pathfinding problem does not actually require full exploration of the state diagram if quasi-optimal solutions are acceptable. Such solutions can be found through local search, which only requires enumeration of a smaller set of states that are transiently reached, plus a slightly larger set of states that are explored and discarded. Approximate algorithms such as Monte Carlo Tree Search (MCTS) (Browne *et al.* 2012) can be used for larger problems by focusing computation only on promising state and action sequences. Moreover, MCTS can handle stochastic transitions, as well as uncertainty in observations of states when the problem is formulated as a partially observable Markov decision process (Katt *et al.* 2017; Lim *et al.* 2021, 2022).

We also assumed that navigation problems involve a single desired state. However in realistic use cases more diffuse targets may exist, e.g. any state with high richness, or where a

**16**

certain species is present, or where mean trait composition is within a certain range. A* cannot handle this scenario, but MCTS can.

In addition, we assumed that the costs of each action are constant by type. However, $\epsilon$-deleting one species might be more costly than for another, either because the time or effort required is high or may depend on whether a third species is also present. Or the costs of different actions may also not be known in advance. Similarly, we assumed that the magnitude of actions ($\epsilon$) is constant. Based on our computational experiments, variation in action costs seems unlikely to substantially impact navigation, whereas variation in action magnitude does, with larger $\epsilon$ enabling more shortcuts. MCTS could also be used to probabilistically identify navigation solutions when costs are unknown or variable (Deglurkar *et al.* 2021).

Last, we assumed that there are no feedbacks among the environment and species, e.g. depletion of limiting resources affecting competition (Tilman 1982). These effects would shift the identity of and relationships among fixed points. Including them is possible if the environment variables can be treated as state variables, which would require some modification of the current implementation.

Trajectories do not necessarily reach fixed points in other models, and could instead reach other attractors like limit cycles. Additionally, multiple fixed points for each combination of species could exist, meaning that the value of $\epsilon$ would take a larger role in determining which basin of attraction was reached. Both scenarios would increase the cardinality of the state space and action space. However, if 'states' and 'actions' can still be defined, then a discretized state diagram can still be constructed.

Safe navigation is also a priority for applications. Navigation should avoid certain states if they are unethical to create, or if their creation would have negative ecological consequences (Aswani *et al.* 2013; Mohseni *et al.* 2021). Notably many paths discovered by our approach transiently put the community into higher-richness states that include novel species (e.g. orange-colored states in **Figure 2**). This strategy may have substantial risk if those novel species escape due to mechanisms not included in the dynamical model. Adding safety constraints could strongly influence reachability of desired states (Bansal & Tomlin 2020) and require algorithms beyond our current implementation.

**Implications for community assembly**

State diagrams provide potential linkages to community assembly, under the assumption that the invasion of new species is infrequent relative to the dynamics. The invasion graph (Hofbauer & Schreiber 2022) is the subgraph of the state diagram comprising only actions that include a single addition and then a wait action (all richness-increasing 'ladders'; **Figure S8**). These actions, and the states they connect, enumerate the most complex communities that can be reached via sequential single invasions. Notably, most states cannot be reached this way; they instead require more complex actions present in the full state diagram (e.g. direct paths involving multiple simultaneous additions and then a wait; or shortcut paths involving multiple wait actions). Conversely, one can also conceptualize an 'un-invasion' graph, which is the subgraph of the state diagram comprising the wait actions linking transient states to fixed point states with no environment change (all richness-decreasing 'snakes'; **Figure S9**). These actions, and the states they connect, enumerate the possible paths by which transiently-reached communities can decay into stable communities. The un-invasion and invasion graphs have non-trivial structures that may be useful for describing community assembly/dis-assembly pathways. We have not yet investigated the general properties of these subgraphs, but see (Hang-Kwang & Pimm 1993; Almaraz *et al.* 2022; Hofbauer & Schreiber 2022).

Second, state diagrams may also help understand priority effects, i.e. order-dependent community assembly (Fukami 2015). This is because repeatedly taking single actions and then waiting potentially has outcomes that depend on the order of operations; more strongly, taking multiple actions at the same time and then waiting may have different consequences than taking each action in sequence. We did not systematically explore such order dependence, but see (Serván *et al.* 2018).

Third, some states may be harder to reach than others, both in community assembly and in navigation. States that have no incident paths are impossible to reach except by brute-force assembly, while those that have very few outgoing paths (especially involving shortcuts) are potentially less likely to reach by chance. These states are related to the 'holes' described by Angulo *et al.* (2021). Initial states with very few outgoing paths are potentially less likely to change state by chance. States that are only reached by 'ladders' may be more easily built up from lower richness states, while states that are only reached by 'snakes' may be more easily broken down from higher richness states. In support of this idea, species combinations most likely to persist under environmental perturbation are more frequent (Medeiros *et al.* 2021). There may also be 'game changing' species (Deng *et al.* 2021) that are disproportionately important for shaping the properties of the state diagram, both in terms of the prevalence and identity of feasible and stable states, as well as the prevalence and identity of shortcuts.

**Conclusion**

State diagrams may be useful for solving applied navigation problems and understanding community assembly. Our current work is limited by its focus on numerical simulation for a single dynamical model. Adapting coexistence theory (Levine *et al.* 2017; Saavedra *et al.* 2017; Gibbs *et al.* 2022) to make general predictions about state diagram topology may be fruitful. Additionally, experimental validation of navigation predictions for community ecology has been absent except in a few microbial (Clark *et al.* 2021; Baranwal *et al.* 2022) and insect (Desharnais *et al.* 2001) cases. Validation is a priority next step for making progress towards real-world applications.

**Acknowledgments**

We are grateful to our colleagues whose public datasets made these empirical analyses possible. Daniel Maynard ran additional analyses to provide us with empirical parameter sets. David Ackerly, Carl Boettiger, Sampada Deglurkar, Shankar Deka, and several anonymous reviewers provided feedback on the manuscript. Courtenay Ray and Pierre Gaüzère helped with initial idea generation. ZS and CT were supported by a DARPA Assured Autonomy grant (FA8750-18-C-0101), the SRC CONIX program (2018-JU-2779), and the NSF Frontiers program (CPS-1545126). ZS was also supported by the University of Colorado Boulder. ML was supported by the NSF Graduate Research Fellowship Program (DGE-1752814, DGE-2146752). BB was supported by the NSF Rules of Life program (DEB-2025282). Any opinions, findings, and conclusions or recommendations expressed in this material are those of the authors and do not necessarily reflect the views of any sponsoring organizations.

**Tables**

**Table 1.** Summary of properties for empirical datasets used in this study. Abundance values are summarized across all assemblages and then across all experimental conditions. The number of edges in the state diagram are summarized across all $\epsilon$ values.

| Dataset | Number of species (n) | Number of environments (m) | Number of edges in state diagram (mean, s.d.) | Proportion of states feasible and stable | Abundance (grand mean, grand s.d.) |
|---|---|---|---|---|---|
| Ciliate | 5 | 1 | 27 ± 2 | 0.25 | 2.75 ± 1.48 |
| Ciliate+environment3 | 5 | 3 | 2851 ± 1562 | 0.9 | 0.07 ± 0.02 |
| Ciliate+environment5 | 5 | 5 | 5667 ± 2820 | 0.74 | 0.61 ± 1.26 |
| Human gut | 12 | 1 | 7132 ± 219 | 0.05 | 2.08 ± 8.35 |
| Mouse gut | 11 | 1 | 22065 ± 1590 | 0.24 | 6.1 ± 46.83 |
| Protist | 11 | 1 | 408 ± 39 | 0.02 | 8.29 ± 65.82 |

**Figures**

**Figure 1. (a)** Navigation is the problem of discovering sequences of actions that shift a community from an initial state (purple circle) to a desired state (green circle) while not unnecessarily visiting other states (gray circles). A direct path (black arrow) involves taking a single low-cost action. A shortcut path (gray arrows connecting orange circles) involves taking several low-cost actions, and represents a 'work-with-nature' solution. A brute-force solution (red arrows) involves deleting all individuals of all undesired species and adding many individuals of all desired species. **(b)** Navigation is loosely analogous to playing 'Snakes and Ladders'. In this game, players transition between squares (states) through sequential movement (actions) that either progress along the board (direct path) or jump around via snakes or ladders (shortcut paths). This game board is from India, Ajmer district, circa 1815. Ashmolean Museum collection EA2007.2, reproduced with permission.[1]

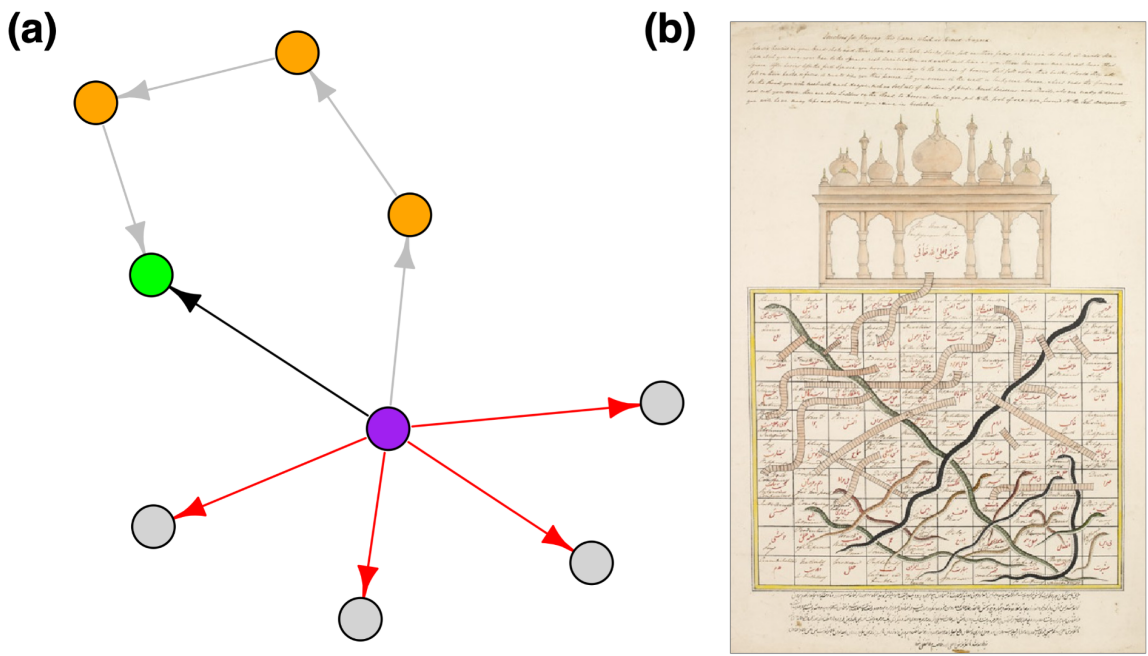

---

[1] A reproduction license is available and will be purchased from the museum by the author upon manuscript acceptance (pers. comm. Ashmolean Museum, May 27, 2021)

**Figure 2.** Example state diagrams for all datasets. Circles represent fixed point states and are colored green if feasible and stable (i.e. possible navigation target), and gray if not. Orange circles indicate transient states that have higher richness than their pre-action state. States are arranged by richness on the y-axis, with the empty state at bottom and the maximum richness state at top. Arrows indicate actions; redder arrows are primarily deletions, while bluer arrows are primarily additions, and intermediate colors indicate mixtures of both additions and deletions; arrow thickness indicates inverse action cost (thicker = lower cost). Panels (b) and (c) indicate cases where there are multiple environments. For visual presentation, environment-changing actions are not separately colored, and states are not ordered by environment (this is why there is more than one state shown at minimum/maximum richness). Visualizations are for $C_{\epsilon-addition} = 1\text{-}, C_{\epsilon-deletion} = 1, C_{environment} = 1, C_{wait} = 0.1$, and $\epsilon = 0.1$. See **Figure S8** for the 'ladder' path subset and **Figure S9** for the 'snake' path subset.

**(a) Ciliate** 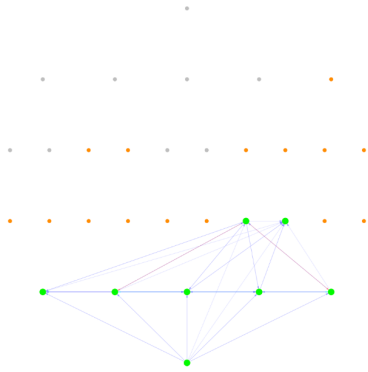 **(b) Ciliate+environment3** 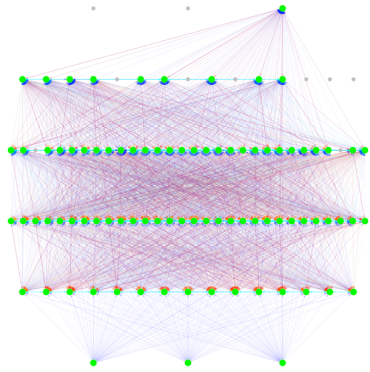 **(c) Ciliate+environment5** 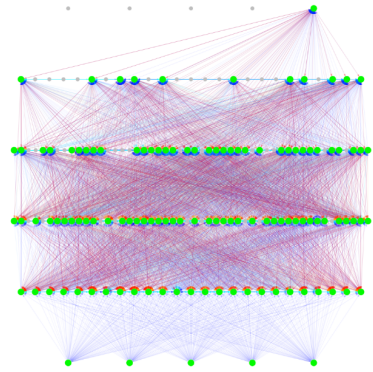

**(d) Human gut** 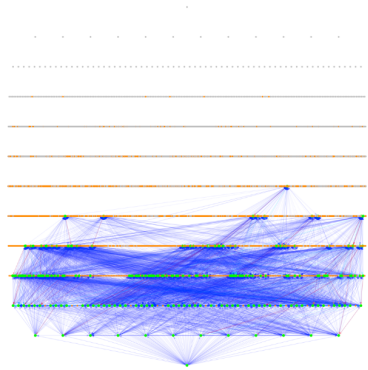 **(e) Mouse gut** 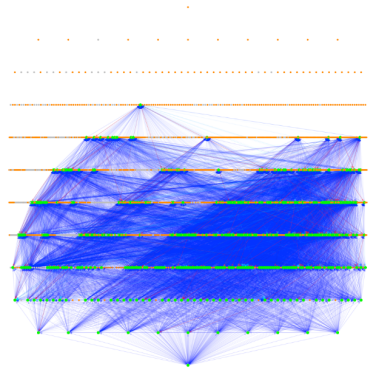 **(f) Protist** 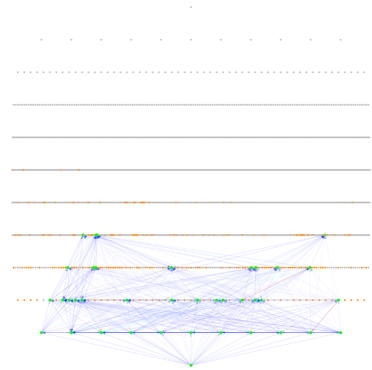

**Figure 3. (a)** Probability that non-brute-force navigation is possible between two randomly selected feasible and stable states. **(b)** Probability that a shortcut path exists between two randomly selected states, given that navigation is possible. Bars indicate different datasets and are colored by ϵ. Error bars in panel **(b)** indicate standard deviations across assumed costs $C_q$; no error bars are shown in **(a)** because costs do not influence estimates.

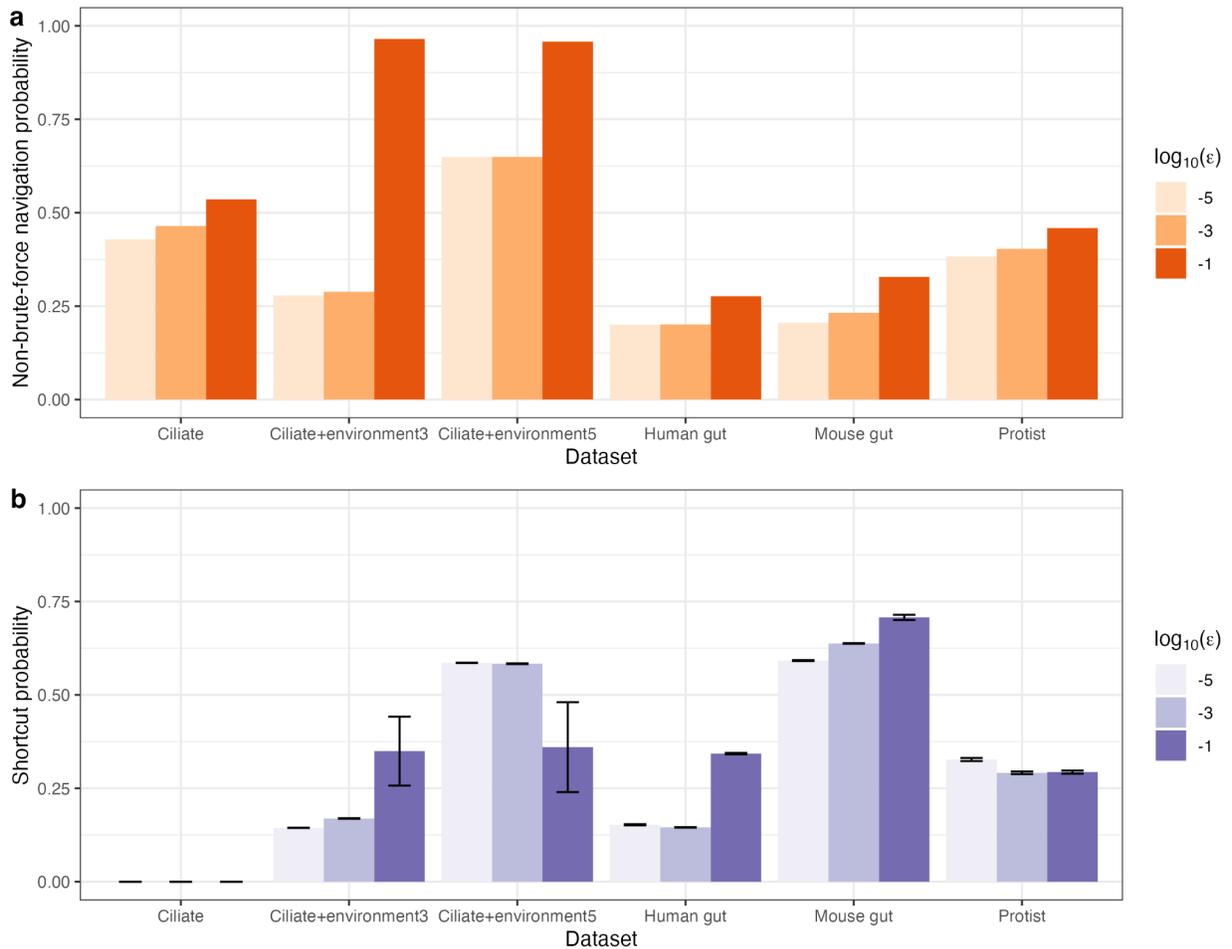

**Figure 4.** Example shortcut paths for **(a-b)** completely removing the pathogen *Clostridium difficile* in the mouse gut dataset, and **(c-d)** reducing species richness via environmental change in the ciliate+environment5 dataset. Each panel indicates states connected by sequential actions on the x-axis, ordered by richness on the y-axis. Green boxes indicate feasible and stable states, with the initial state on the left and the desired state on the right. White boxes indicate actions, with ϵ-additions as blue '+', ϵ-deletions as red '-', environment changes as orange '*', and waits as gray '.'. Visualizations are for $C_{\epsilon-addition} = 1$, $C_{\epsilon-deletion} = 10$, $C_{environment} = 1$, $C_{wait} = 0.1$, and $\epsilon = 0.1$ (i.e. 10× more costly to ϵ-delete than ϵ-add) Taxon names are in **Table S1**.

**(a)** 1: One species is introduced at low density and another is given a small negative abundance perturbation, causing the establishment of one species and the competitive exclusion of three others. 2: Three species are introduced at low density, causing the competitive exclusion of *C. difficile*. 3: Two species are introduced at low density, yielding competitive exclusion of one species and coexistence of four species in the desired state. **(b)** 1: Two species are introduced at low density, causing one competitive exclusion. 2: One species is introduced at low density, causing two competitive exclusions. 3: Three species are introduced at low density, causing the competitive exclusion of *C. difficile* and two other species. 4: Two species are introduced at low density, yielding competitive exclusion of three species and coexistence of two species in the desired state. **(c)** 1: The environment is warmed, causing competitive exclusion of two species. 2: One species is introduced at low density and the environment is cooled, causing coexistence of two species in the desired state. **(d)** Similar to **(c)**.

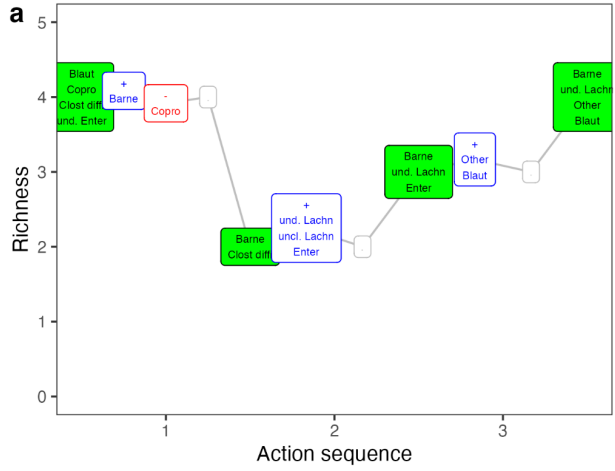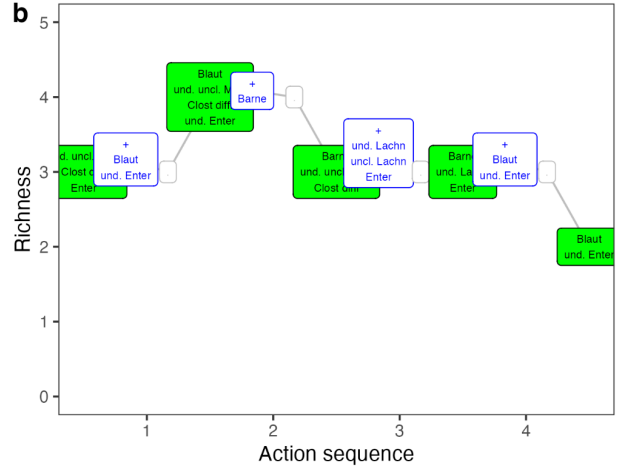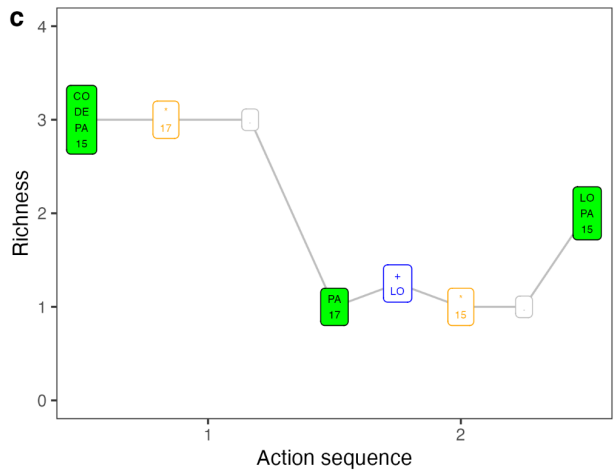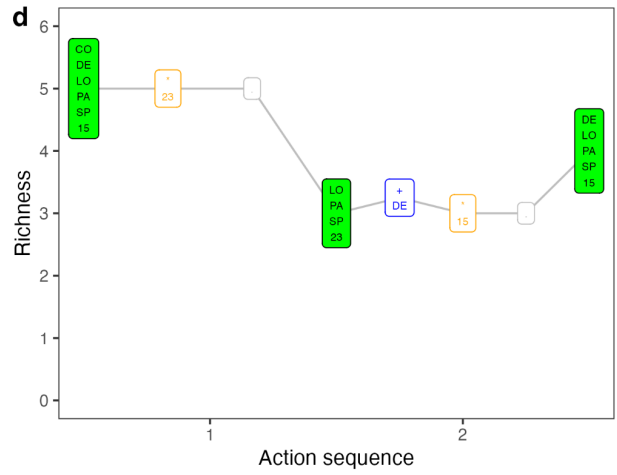

**Figure 5.** Partial dependence plots indicating the effect of each individual predictor on the probability of navigation yielding a brute-force solution, direct path, or shortcut path.

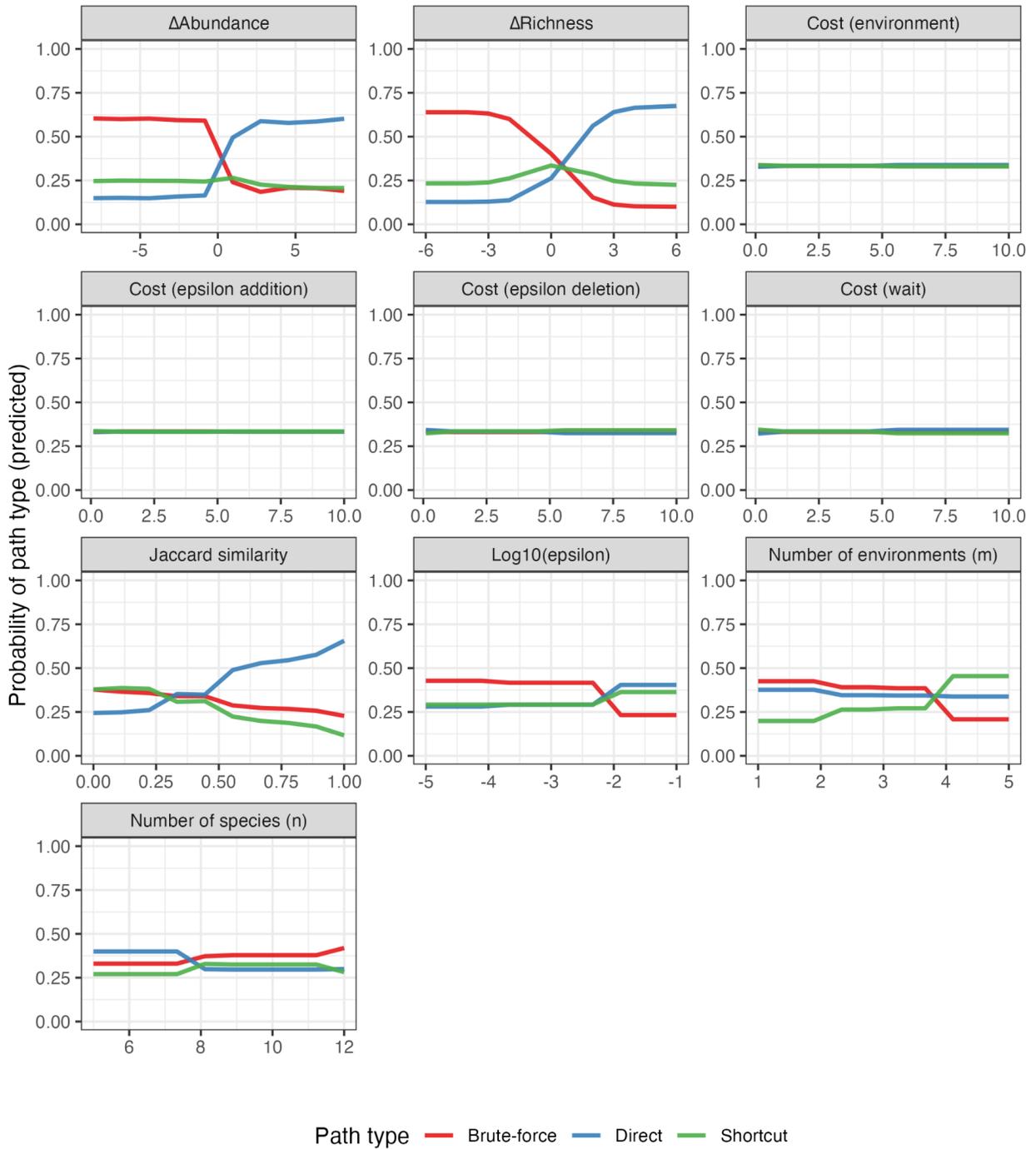

**Supporting Information**

**Figure S1.** Distribution of actions within edges for the state diagrams shown in **Figure 2**. Panels indicate the number of ϵ-additions (x-axis), ϵ-deletions (y-axis), and environmental change actions (facets) comprising each edge in the state diagram.

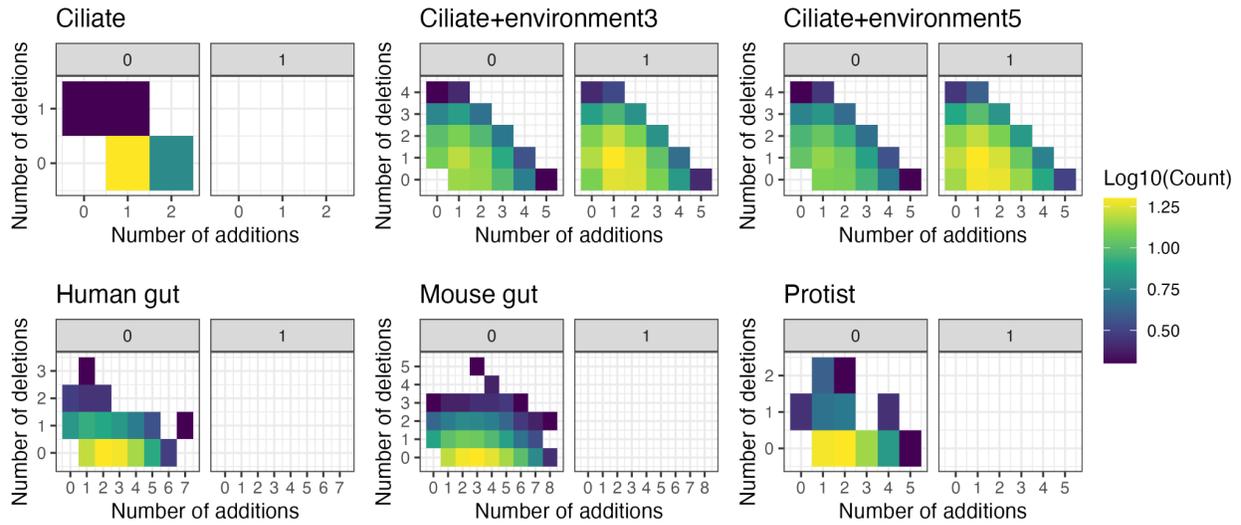

**Figure S2.** State diagram properties vary with ϵ. Columns indicate datasets; rows indicate low to high values of ϵ. All labels are as in **Figure 2**.

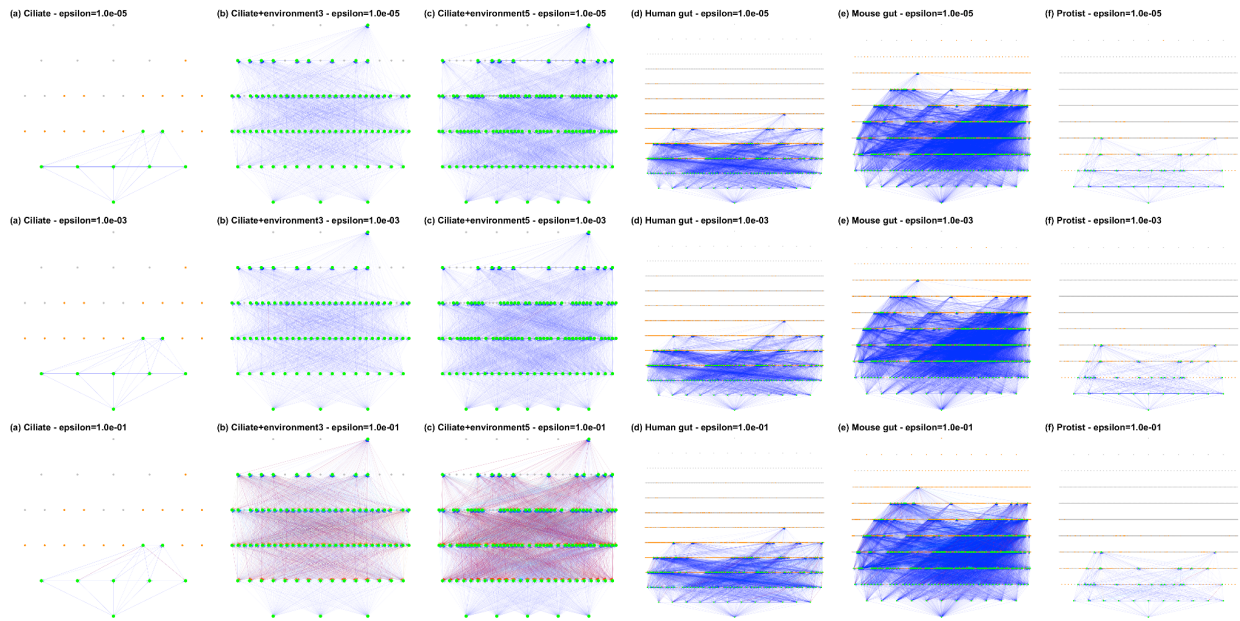

**Figure S3.** Most-visited intermediate states. For each dataset and A* experiment we identified the intermediate states that were most commonly visited in navigation among all state pairs. We then rank-ordered these states by their prevalence within each dataset and A* experiment. Panels show states on the y-axis and experiments on the x-axis. Top-5 common states are colored in red; all others in blue. Panels are faceted by $\epsilon$. Columns within each panel indicate different experiments, i.e. different combinations of costs $C_i$. The prevalence of apparent horizontal red lines within each each facet indicates that the identity of most-visited intermediate states is sometimes not strongly dependent on $C_q$.

### Ciliate
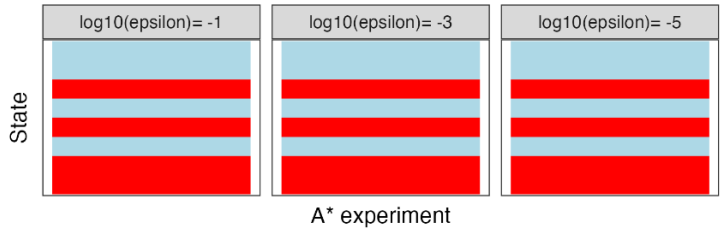

### Ciliate+environment3
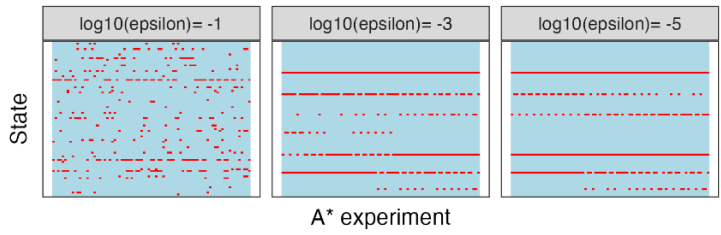

### Ciliate+environment5
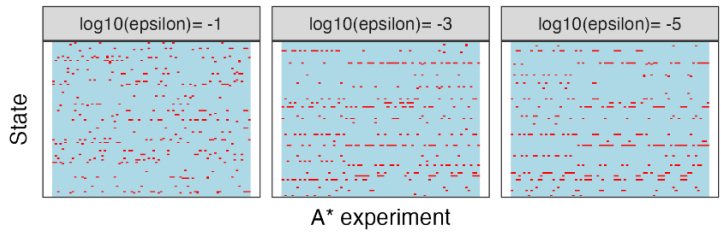

### Human gut
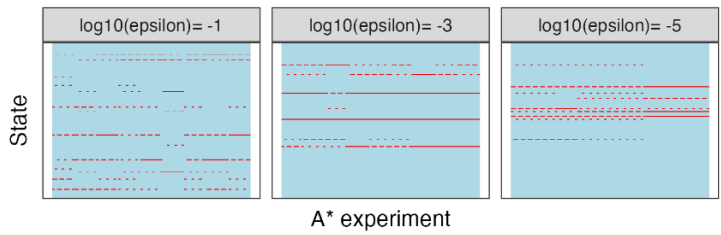

### Mouse gut
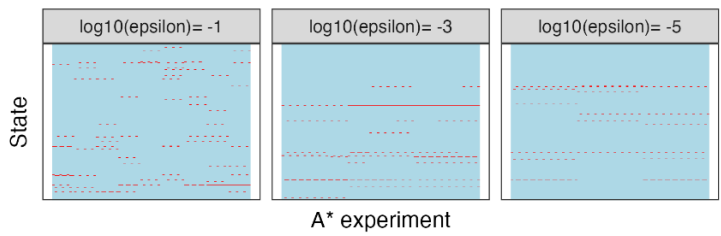

### Protist
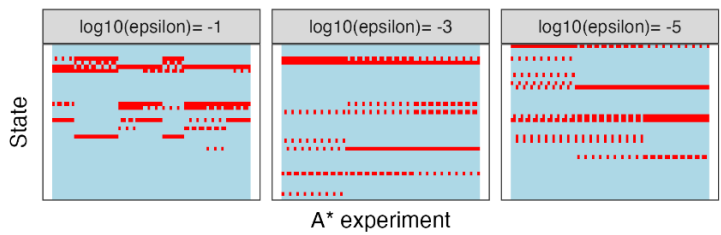

**Figure S4.** Summary of network topology for state diagrams for different datasets (rows) and $\epsilon$ values (columns). Facet points indicate in- and out-degree for each state, and are colored by the number of intermediate visits. Gray points indicate states that are not reachable by non-brute-force navigation. Data are shown for a case where all $C_q = 1$ for all action types; results do not vary strongly with $C_q$ (not shown).

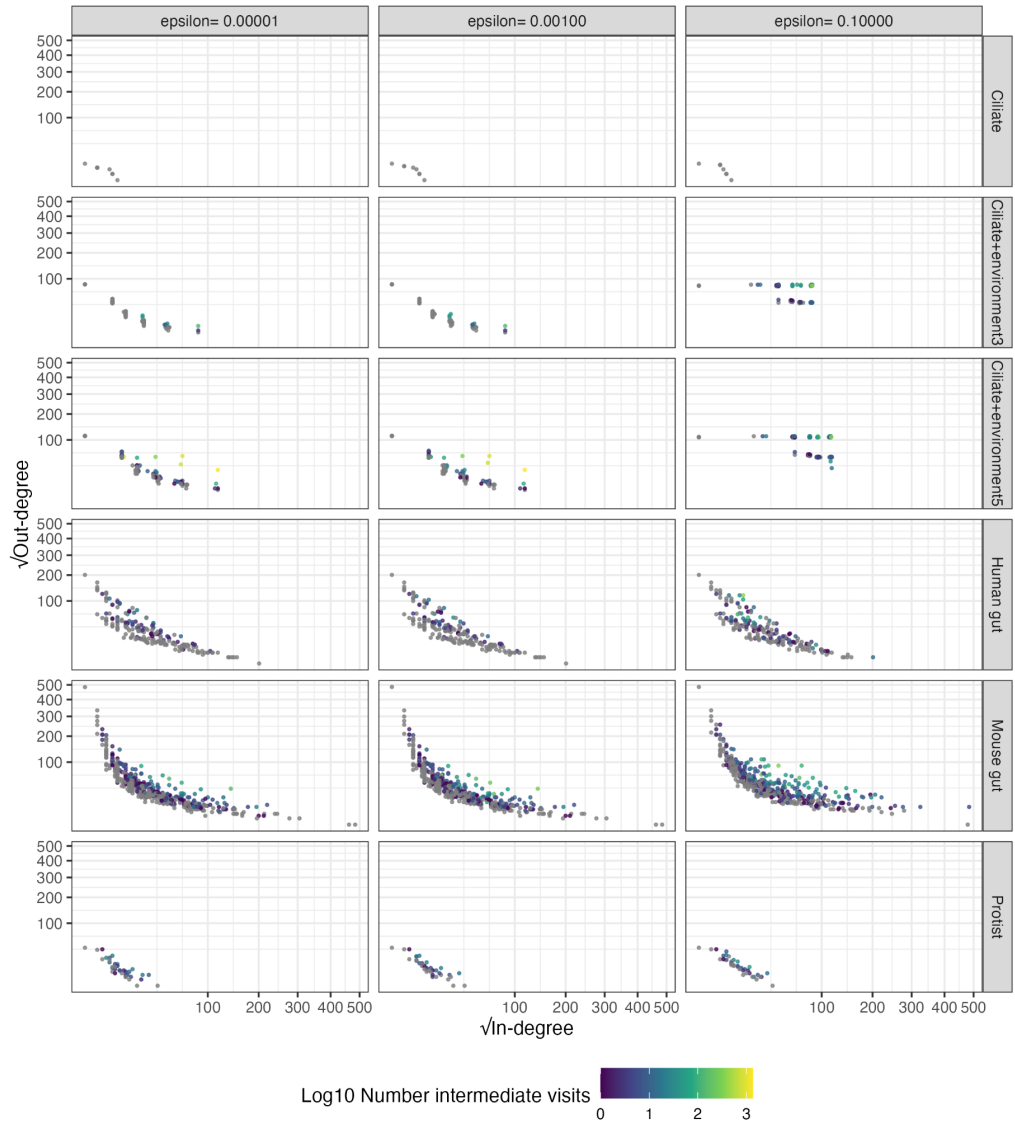

**Figure S5.** Distribution of path lengths among shortcuts. Panels are faceted by dataset; line color indicates $\epsilon$; multiple lines represent variation due to assumed costs $C_q$.

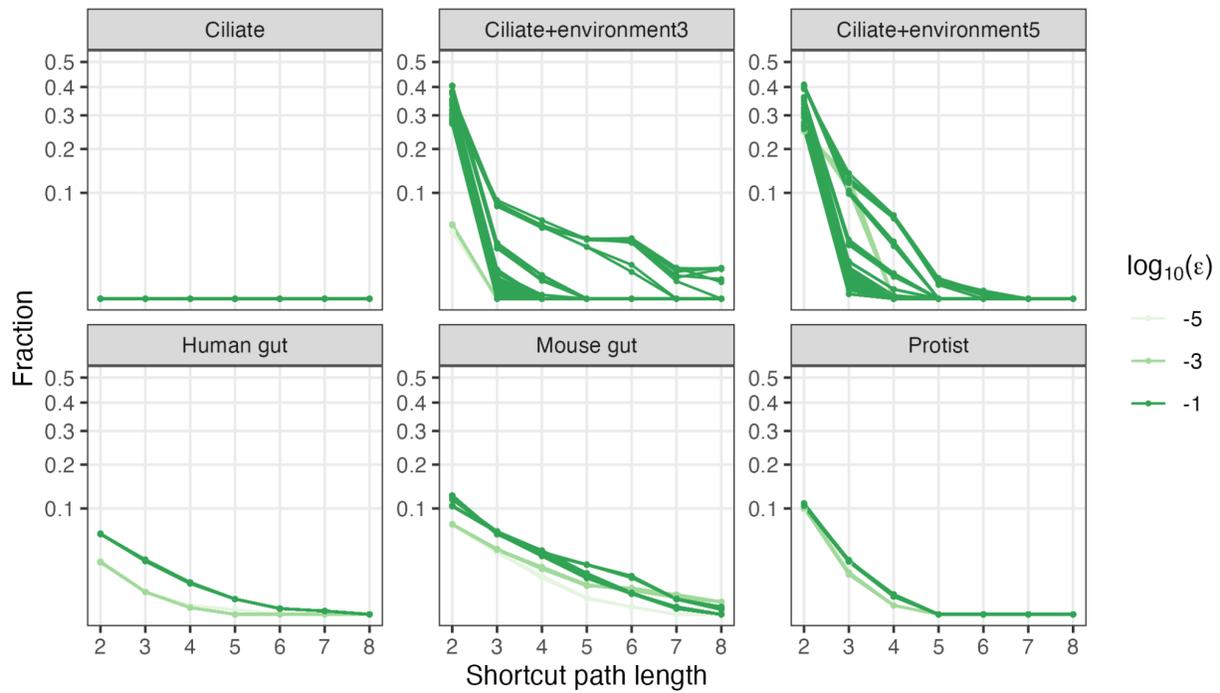

**Figure S6.** Variable importance plot for the random forest model shown in **Figure 5**.

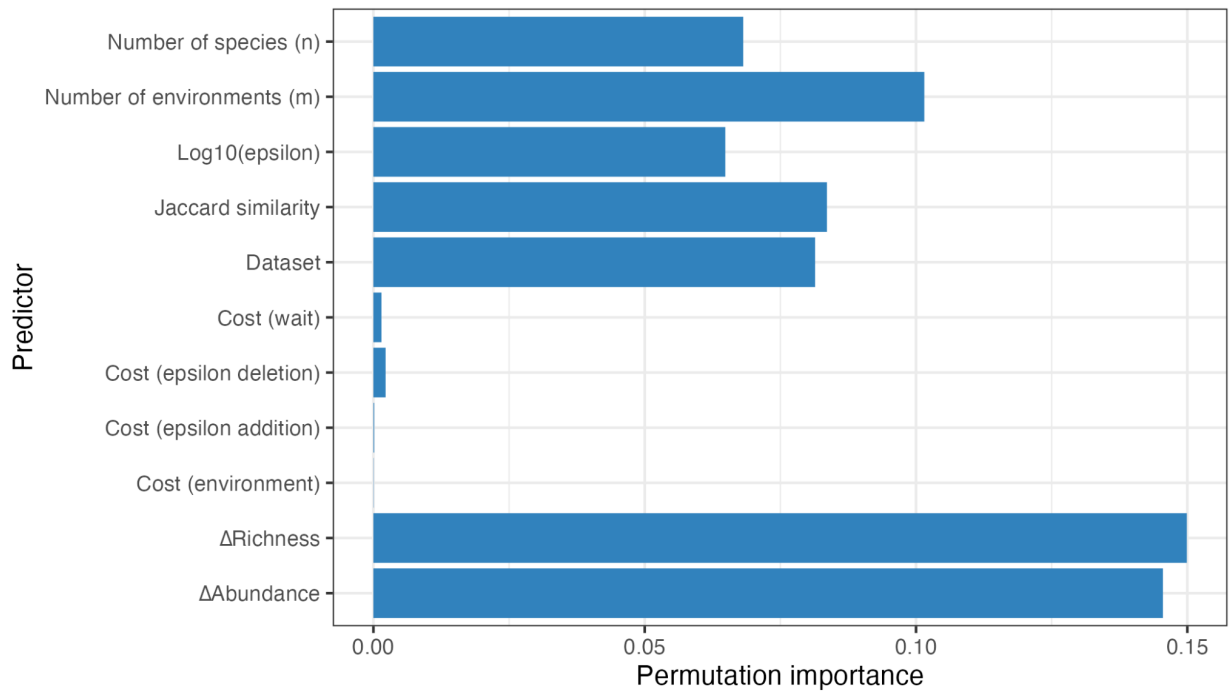

**Figure S7.** Partial dependence plots indicate the effect of ΔAbundance and ΔRichness on the probability of navigation yielding no path (brute-force solution), a direct path, or a shortcut path. Δs are defined as desired state values minus initial state values.

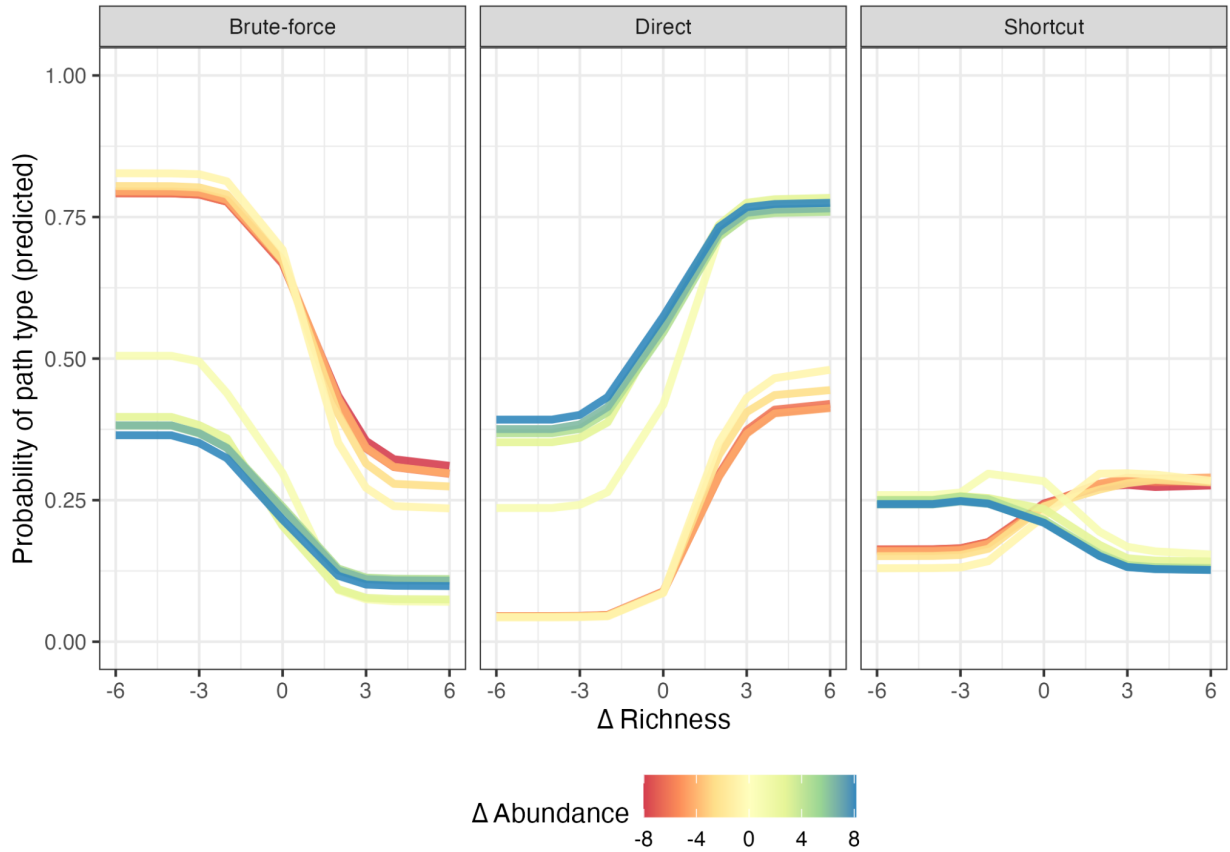

**Figure S8.** Invasion graphs for all datasets. Notation and parameters are the same as in **Figure 2**, except that transient states are not colored orange.

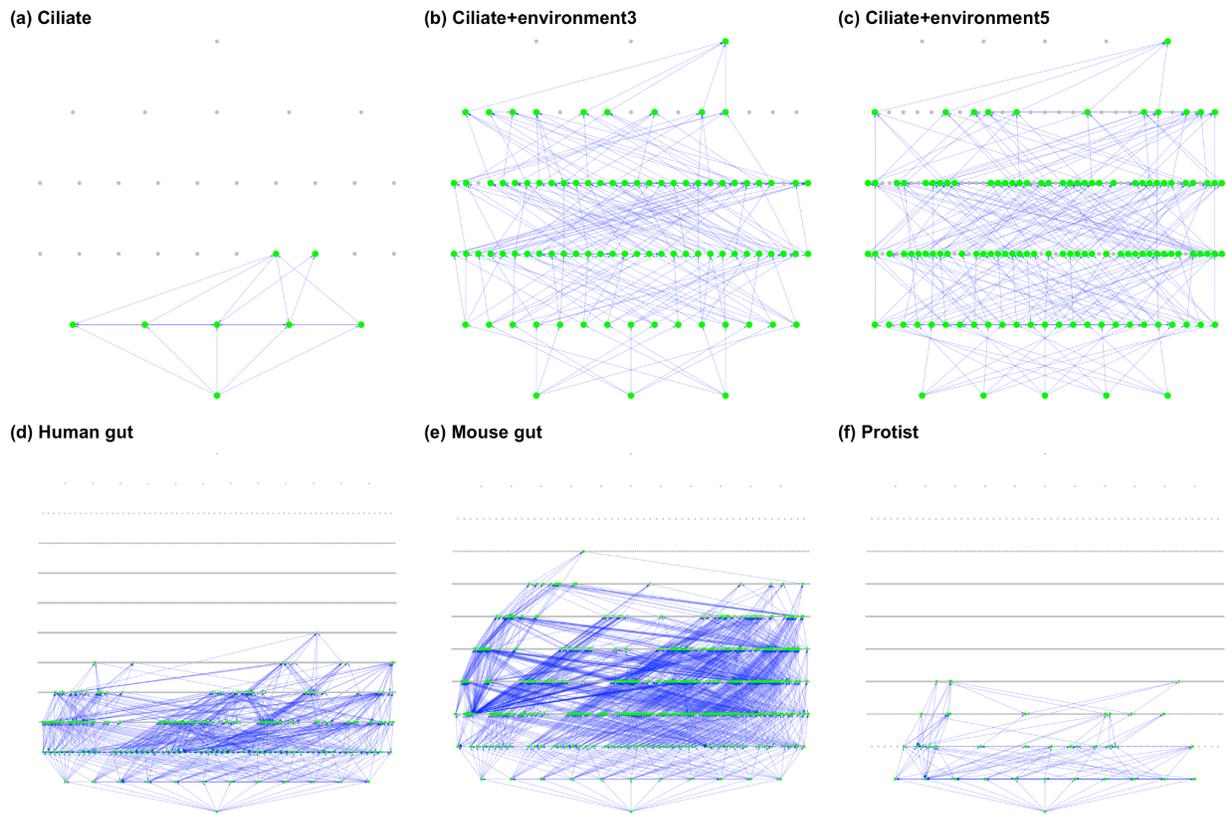

**Figure S9.** Un-invasion graphs for all datasets. Notation and parameters are the same as in **Figure 2**.

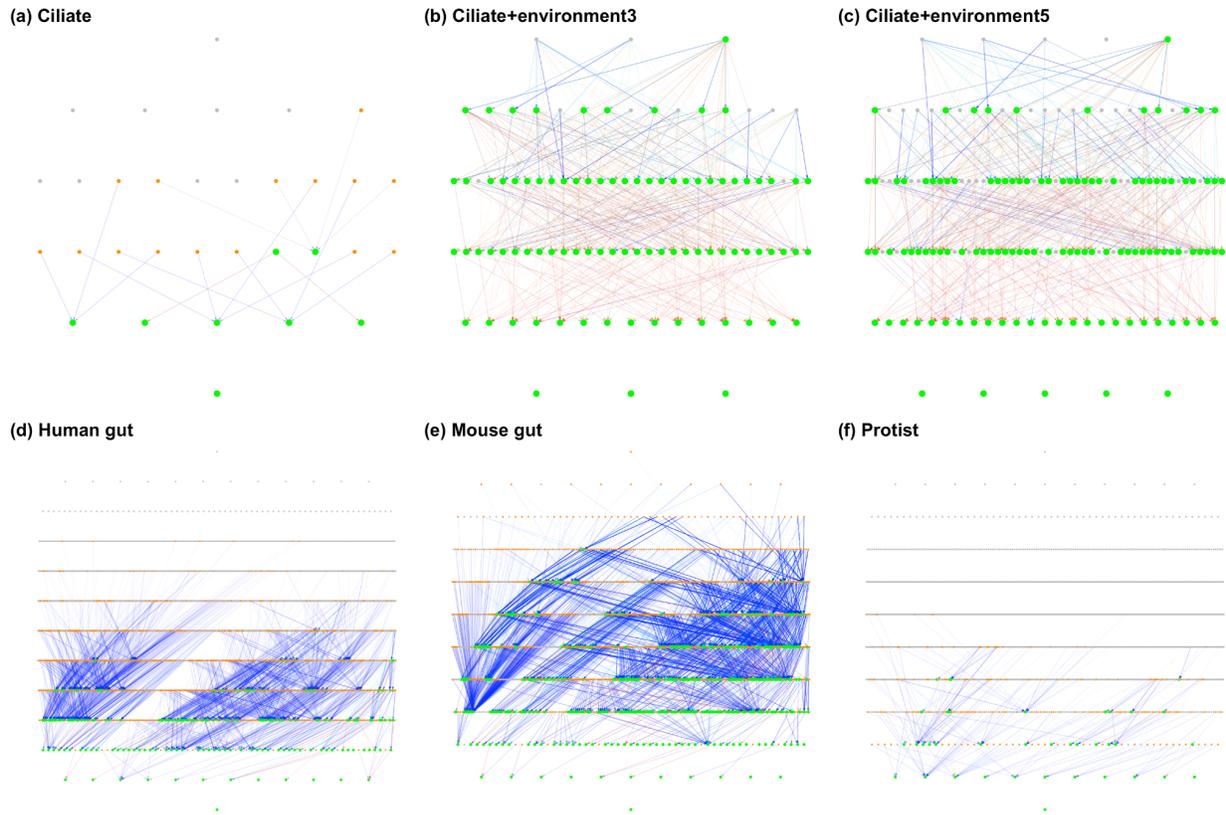

**Table S1.** Taxon names within each dataset. Column 'Dataset' indicates the dataset name as used in the main text. Column 'Filename in code' indicates the dataset abbreviation used in the code repository. Column 'Species number in code' indicates the numeric species abbreviation used in the code repository. Column 'Taxon name in code' indicates the taxon abbreviation in code. Column 'Taxon name' indicates the scientific name of the taxon.

| Dataset | Filename in code | Species number in code | Taxon name in code | Taxon name |
|---|---|---|---|---|
| Ciliate | Maynard | 1 | CO | *Colpidium striatum* |
| Ciliate | Maynard | 2 | DE | *Dexiostoma campylum* |
| Ciliate | Maynard | 3 | LO | *Loxocephalus sp.* |
| Ciliate | Maynard | 4 | PA | *Paramecium caudatum* |
| Ciliate | Maynard | 5 | SP | *Spirostomum teres* |
| Ciliate+environment3 | Maynard15-19-23 | 1 | CO | *Colpidium striatum* |
| Ciliate+environment3 | Maynard15-19-23 | 2 | DE | *Dexiostoma campylum* |
| Ciliate+environment3 | Maynard15-19-23 | 3 | LO | *Loxocephalus sp.* |
| Ciliate+environment3 | Maynard15-19-23 | 4 | PA | *Paramecium caudatum* |
| Ciliate+environment3 | Maynard15-19-23 | 5 | SP | *Spirostomum teres* |
| Ciliate+environment5 | Maynard15-17-19-21-23 | 1 | CO | *Colpidium striatum* |

| Ciliate+environment5 | Maynard15-17-19-21-23 | 2 | DE | *Dexiostoma campylum* |
|---|---|---|---|---|
| Ciliate+environment5 | Maynard15-17-19-21-23 | 3 | LO | *Loxocephalus sp.* |
| Ciliate+environment5 | Maynard15-17-19-21-23 | 4 | PA | *Paramecium caudatum* |
| Ciliate+environment5 | Maynard15-17-19-21-23 | 5 | SP | *Spirostomum teres* |
| Human gut | Venturelli | 1 | BH | *Blautia hydrogenotrophica* |
| Human gut | Venturelli | 2 | CA | *Collinsella aerofaciens* |
| Human gut | Venturelli | 3 | BU | *Bacteroides uniformis* |
| Human gut | Venturelli | 4 | PC | *Prevotella copri* |
| Human gut | Venturelli | 5 | BO | *Bacteroides ovatus* |
| Human gut | Venturelli | 6 | BV | *Bacteroides vulgatus* |
| Human gut | Venturelli | 7 | BT | *Bacteroides thetaiotaomicron* |
| Human gut | Venturelli | 8 | EL | *Eggerthella lenta* |
| Human gut | Venturelli | 9 | FP | *Faecalibacterium prausnitzii* |
| Human gut | Venturelli | 10 | CH | *Clostridium hiranonis* |
| Human gut | Venturelli | 11 | DP | *Desulfovibrio piger* |
| Human gut | Venturelli | 12 | ER | *Eubacterium rectale* |
| Mouse gut | Bucci | 1 | Barne | *Barnesiella* |
| Mouse gut | Bucci | 2 | und. Lachn | und. Lachnospiraceae |
| Mouse gut | Bucci | 3 | uncl. Lachn | uncl. Lachnospiraceae |
| Mouse gut | Bucci | 4 | Other | Other |
| Mouse gut | Bucci | 5 | Blaut | *Blautia* |
| Mouse gut | Bucci | 6 | und. uncl. Molli | und. uncl. Mollicutes |
| Mouse gut | Bucci | 7 | Akker | *Akkermansia* |
| Mouse gut | Bucci | 8 | Copro | *Coprobacillus* |
| Mouse gut | Bucci | 9 | Clost diffi | *Clostridium difficile* |
| Mouse gut | Bucci | 10 | Enter | *Enterococcus* |
| Mouse gut | Bucci | 11 | und. Enter | und. Enterobacteriaceae |
| Protist | Carrara | 1 | Chi | *Chilomonas* sp. |
| Protist | Carrara | 2 | Cyc | *Cyclidium* sp. |
| Protist | Carrara | 3 | Tet | *Tetrahymena* sp. |
| Protist | Carrara | 4 | Dex | *Dexiostoma* sp. |
| Protist | Carrara | 5 | Col | *Colpidium* sp. |
| Protist | Carrara | 6 | Pau | *Paramecium aurelia* |
| Protist | Carrara | 7 | Cep | *Cephalodella* sp. |
| Protist | Carrara | 8 | Spi | *Spirostomum* sp. |
| Protist | Carrara | 9 | Eug | *Euglena gracilis* |
| Protist | Carrara | 10 | Eup | *Euplotes aediculatus* |
| Protist | Carrara | 11 | Pbu | *Paramecium bursaria* |

**Table S2.** Environment names within each dataset. Column 'Dataset' indicates the dataset name as used in the main text. Column 'Filename in code' indicates the dataset abbreviation used in the code repository. Column 'Environment number in code' indicates the numeric environment abbreviation used in the code repository. Column 'Environment name' indicates the biological name of the environment.

| Dataset | Filename in code | Environment number in code | Environment name |
|---|---|---|---|
| Ciliate | Maynard | 1 | 17 °C |
| Ciliate+environment3 | Maynard15-19-23 | 1 | 15 °C |
| Ciliate+environment3 | Maynard15-19-23 | 2 | 19 °C |
| Ciliate+environment3 | Maynard15-19-23 | 3 | 23 °C |
| Ciliate+environment5 | Maynard15-17-19-21-23 | 1 | 15 °C |
| Ciliate+environment5 | Maynard15-17-19-21-23 | 2 | 17 °C |
| Ciliate+environment5 | Maynard15-17-19-21-23 | 3 | 19 °C |
| Ciliate+environment5 | Maynard15-17-19-21-23 | 4 | 21 °C |
| Ciliate+environment5 | Maynard15-17-19-21-23 | 5 | 23 °C |
| Human gut | Venturelli | 1 | - |
| Mouse gut | Bucci | 1 | - |
| Protist | Carrara | 1 | - |